\documentclass[journal=jpclcd,layout=manuscript,manuscript=letter]{achemso}
\setkeys{acs}{doi = true}

\usepackage[version=3]{mhchem} 
\usepackage{booktabs}
\usepackage{graphicx}
\usepackage[table,xcdraw]{xcolor}
\usepackage{chemformula} 
\usepackage[T1]{fontenc} 
\usepackage{lscape}

\author{Golokesh Santra}
\affiliation[Weizmann Institute of Science ]{Department of Molecular Chemistry and Materials Science, Weizmann Institute of Science, 7610001 Re\d{h}ovot, Israel}
\alsoaffiliation[MPI Kohlenforschung]{Present address: Max-Planckinstitut f\"ur Kohlenforschung, Kaiser-Wilhelm-Platz 1, 45470 M\"ulheim an der Ruhr, Germany}
\altaffiliation{Equally contributing first author}
\author{Margarita Shepelenko}
\altaffiliation{Equally contributing first author}
\author{Emmanouil Semidalas}
\author{Jan M. L. Martin}

\email{gershom@weizmann.ac.il}
\phone{+972-8-9342533}
\fax{+972-8-9343029}
\affiliation[Weizmann Institute of Science ]
{Department of Molecular Chemistry and Materials Science, Weizmann Institute of Science, 7610001 Re\d{h}ovot, Israel}

\usepackage{xcolor}
\usepackage{etoolbox}
\usepackage{verbatim}

\title[post-CCSD(T) in water clusters]
  {Is valence CCSD(T) enough for the binding of water clusters? The isomers of (H$_2$O)$_6$ and (H$_2$O)$_{20}$ as a case study}
\begin{document}


\begin{tocentry}
\includegraphics[width=2in]{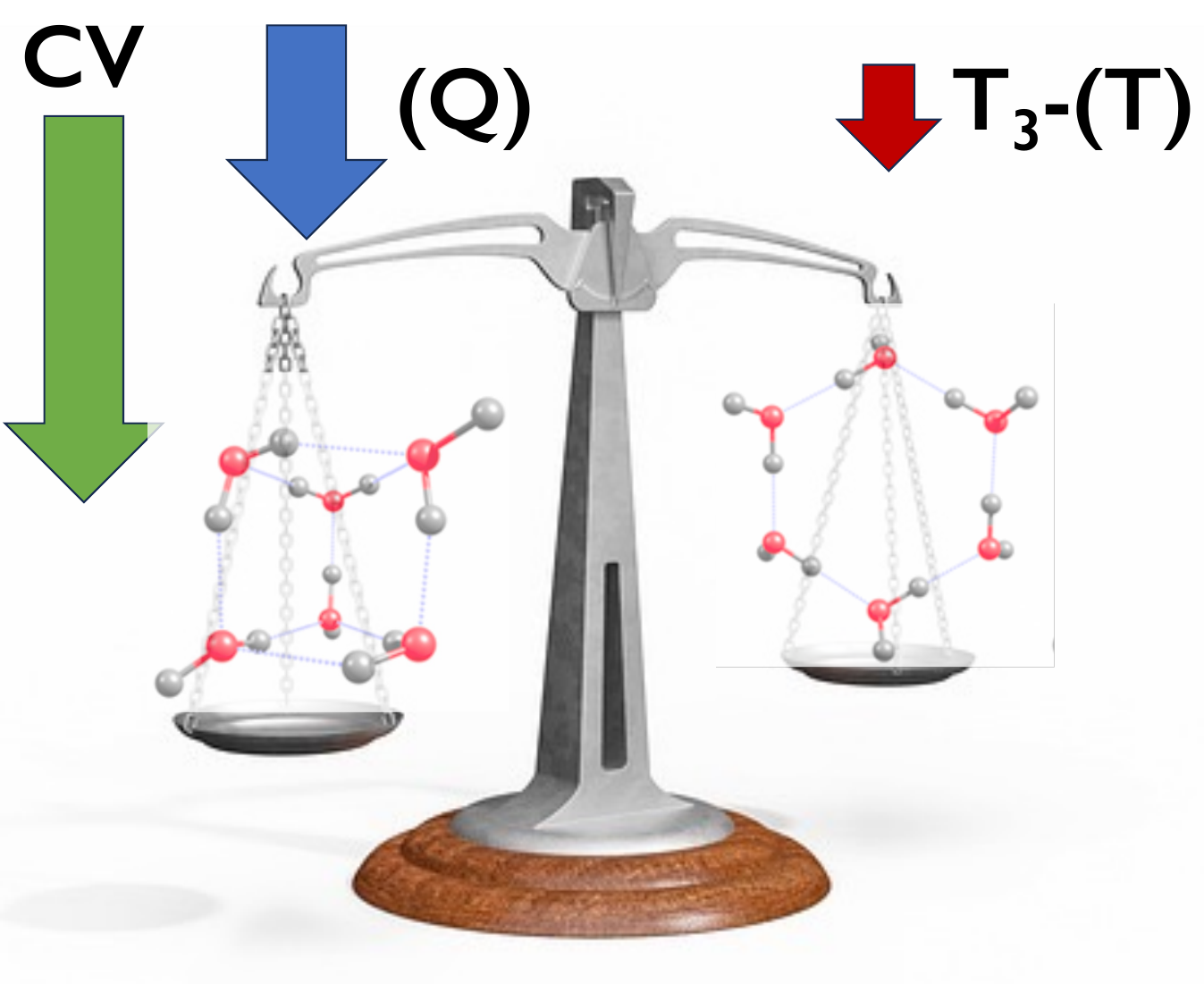}
\end{tocentry}

\begin{abstract}
Benchmark calculations on noncovalent interactions typically exclude correlation effects beyond valence CCSD(T) owing to their steep computational cost scaling. In this work, we consider their importance for water clusters, specifically, eight isomers of (H$_2$O)$_6$ and four Wales-Hodges isomers of (H$_2$O)$_{20}$. Higher order connected triples, $T_3$--(T), reduce dissociation energies of the latter by about 0.4 kcal/mol, but this is more than compensated by an increase of up to 0.85 kcal/mol due to connected quadruple excitations. In general, higher-order correlation effects favor more compact isomers over more `spread-out' ones. We also consider additional small effects for balance: scalar relativistics reduce binding in (H$_2$O)$_{20}$ by ca. --0.4 kcal/mol, which fortuitously is compensated by the ca. 0.55 kcal/mol diagonal Born-Oppenheimer correction. Core-valence correlation has the greatest impact, at ca. 1.3 kcal/mol for the icosamer.
\end{abstract}

\section{Introduction}

Noncovalent interactions are ubiquitous in supramolecular chemistry; chemical biology, e.g., molecular recognition and secondary structure of biomolecules; physical chemistry of the liquid state; and more. Among such interactions, water clusters hold a special position, in attempts to understand the unique physicochemical properties of ``life's solvent" (see the introductions to Refs.\citenum{Paesani2020,XantheasAtlas,Saykally2001,jmlm272,Bowman2021,Tschumper2014,Burke2023}).

The `gold standard' for benchmarking noncovalent interactions has for some time been valence CCSD(T) (coupled cluster with all singles and doubles and quasiperturbative connected triples\cite{Rag89,Wat93}) extrapolated to the complete basis set limit by various stratagems (see, e.g., the very recent review by Karton\cite{Karton2022}). For larger molecules, localized coupled cluster approaches are increasingly employed. 

Owing to the very steep computational cost scaling of higher-order electron correlation methods — for instance, CCSDT\cite{Noga1986} and CCSDT(Q)\cite{mrcc8} will scale as the eighth and ninth (!) power, respectively, of the system size — there have been few efforts to attempt to determine post-CCSD(T) corrections, and these have focused on small species. Boese\cite{Boese2013} considered hydrogen bonding of diatomics, triatomics, and tetratomics; Hobza and coworkers\cite{A24,A24bis} focused on their A24 benchmark; while recently Martin and Karton\cite{jmlm298} investigated the benzene dimer. It is well established in computational thermochemistry (e.g., Refs.\citenum{jmlm200,heat3,jmlm273,Karton2022}) that for molecules dominated by dynamical (rather than static\cite{Hollett2011}) correlation, CCSD(T) is a Pauling point owing to error cancellation between neglect of, usually antibonding, higher-order triple excitations CCSDT -- CCSD(T), and universally bonding connected quadruple excitations.

Application of CCSDT(Q) to problems like the L7 (seven large noncovalent complexes) benchmark\cite{L7} is not realistic at present ---  and yet here, discrepancies between approximate CCSD(T) and quantum Monte Carlo approaches\cite{Hamdani2021} prompt the question whether the error cancellation in CCSD(T) is starting to break down.

One promising system for shedding more light on this question are neutral water clusters (H$_2$O)$_n$, since a many-body expansion(MBE, Ref.\citenum{PolishMBEpaper} and references therein) of their energies converges fairly rapidly\cite{Paesani2016,Herbert2017,jmlm272}, and we saw indications\cite{jmlm272} that the further up the electron correlation `ladder' one climbs, the faster the MBEs for the successive contributions converge. This offers the tantalizing prospect that post-CCSD(T) contributions of fairly large water clusters may be amenable to evaluation through MBE, and that thus insights on the nature of post-CCSD(T) contributions to noncovalent interactions may be obtained. 

We have explored this hypothesis for two systems. The first are the eight isomers of (H$_2$O)$_6$, which has been extensively studied as the smallest cluster that has a three-dimensional (rather than cyclic) global minimum structure. The second are the four (H$_2$O)$_{20}$ isomers that occur in the WATER27 dataset\cite{WATER27}  --- the said isomers represent the four Wales-Hodges\cite{Wales1998} structure classes: dodecahedron, face-cube (a.k.a., box-kite, stacked cubes), face-sharing (stacked pentagons), and edge-shared.

For the sake of perspective, we have also evaluated other contributions that might conceivably affect cluster binding energies in the same energy range, such as core-valence correlation, scalar relativistics, and the diagonal Born-Oppenheimer correction.

\section{Methods}

All CCSD(T) or lower-level calculations reported here were carried out using MOLPRO 2022.3\cite{Molpro2020}, aside from the diagonal Born-Oppenheimer corrections (DBOC), which were obtained using the implementation\cite{Gauss2006} in CFOUR 2.1\cite{Cfour2020}. CCSDT, CCSDT(Q), and CCSDTQ results were likewise obtained using CFOUR (specifically the NCC module\cite{Matthews2015} developed by Devin A. Matthews and coworkers), while CCSDTQ(5) calculations were carried out by means of MRCC\cite{MRCC,mrcc63}. The Weizmann Institute's Faculty of Chemistry HPC facility CHEMFARM was used throughout. As in Ref.\citenum{jmlm272}, structures for the eight (H$_2$O)$_6$ isomers were downloaded from the BEGDB noncovalent interactions database\cite{BEGDB,Temelso2011} and used as is, while those for the four (H$_2$O)$_{20}$ isomers were taken from the WATER27 benchmark (also part of the GMTKN55\cite{GMTKN55}~---~general main-group thermochemistry, kinetics, and noncovalent interactions --- benchmark suite).

Basis sets used for the post-CCSD(T) and DBOC correction are the Dunning correlation consistent cc-pV$n$Z ($n$=D, T, Q) basis sets\cite{Dun89}, as well as their (diffuse function-)augmented variants\cite{Kendall1992} aug-cc-pV$n$Z. For the core-valence calculations, we employed aug-cc-pwCV$n$Z (n=T, Q) basis sets\cite{Peterson2002}. (For a detailed discussion of basis set convergence in this contribution, see Ref.\citenum{jmlm281}.) 

Scalar relativistic corrections were obtained using the 2nd-order Douglas-Kroll-Hess approximation\cite{Reiher2012} with a fully decontracted aug-cc-pVTZ basis set (AVTZuc).

The `brace' notation, e.g., V\{D,T\}Z, refers to extrapolation from cc-pVDZ and cc-pVTZ basis sets. Extrapolation parameters were taken from Table 3 of Karton\cite{Karton2020}. (For a discussion of the equivalence relationships between various two-point extrapolations, see Ref.\citenum{jmlm280}.)

\section{Results and discussion}

Results for the eight (H$_2$O)$_6$ isomers are gathered in Table~\ref{tab:water6}, while those for the four Wales-Hodges (H$_2$O)$_{20}$ structures can be found in Table~\ref{tab:water20}.

The V\{D,T\}Z two-body higher-order triples corrections, $T_3$--(T), are universally antibonding, by about $-0.14\pm$0.01 kcal/mol for the eight (H$_2$O)$_6$ isomers and --0.63$\pm$0.01 kcal/mol for the four (H$_2$O)$_{20}$ isomers. Notably, these contributions are remarkably similar between the different structures. 
In contrast, the 3-body $T_3$-(T) contributions are found to be bonding; the V\{D,T\}Z numbers are a modest 0.02-0.03 kcal/mol for the (H$_2$O)$_6$ isomers, but for (H$_2$O)$_{20}$ this increases to 0.13 kcal/mol for the dodecahedron and $0.18\pm0.01$ kcal/mol for the other structures. 
For (H$_2$O)$_6$, we find the four-body contribution to be altogether insignificant, and the remainder (evaluated as the whole-cluster dissociation energy minus the sum of two-body, three-body, and four-body terms) does not exceed 0.001 kcal/mol. We hence made no attempt to evaluate them for the water icosamers.

\begin{table}[h!]

\caption{Dissociation energy contributions for (H$_2$O)$_{6}$ isomers (kcal/mol)\label{tab:water6}}
\centering
\resizebox{\textwidth}{!}{
\begin{tabular}{lrrrrrrrr}\hline
& prism & cage & book1 & book2 & bag & cyclic chair & boat1 & boat2\\
& Iso1 PR & Iso2 CAG & Iso3 BK1 &Iso4 BK2 & Iso5 BAG & Iso6 CC & Iso7 CB1 & Iso8 CB2 \\
\hline
&\multicolumn{8}{c}{valence CCSD(F12*)/CBS +(T) limit}\\
Ref.\cite{jmlm272}& 48.88 & 48.53 & 48.10 & 47.83 & 47.37 & 46.98 & 45.98 & 45.88\\
\hline
$T_3-$(T) V\{D,T\}Z 2-body& -0.153 & -0.156 & -0.146 & -0.143 & -0.152 & -0.137 & -0.139 & -0.136\\
$T_3-$(T) V\{T,Q\}Z 2-body& -0.143 & & & & & -0.141 \\
$T_3-$(T) VDZ 3-body & 0.014 & 0.014 & 0.009 & 0.010 & 0.010 & 0.004 & 0.004 & 0.004 \\
$T_3-$(T) VTZ(d,p) 3-body & 0.021 & 0.018 & 0.016 & 0.016 & 0.016  & 0.012 & 0.012  & 0.012  \\
$T_3-$(T) VTZ(f,p) 3-body & 0.022 & 0.020 & 0.016 & 0.016 & 0.016 & 0.011 &0.011 &0.012\\
$T_3-$(T) V\{D,T\}Z 3-body & 0.026& 0.022& 0.019& 0.019& 0.019& 0.015& 0.015& 0.016\\
$T_3-$(T) VDZ 4-body & 0.005 & 0.003 & 0.003 & 0.004 & 0.003 & 0.002 & 0.002 & 0.002 \\
$T_3-$(T) VDZ 5+6-body & 0.000 & 0.000 & 0.000 & 0.000 & 0.001 & 0.001 & 0.001 & 0.001\\
\hline
(Q) 2-body &0.191 & 0.184 & 0.154 & 0.155 & 0.158 & 0.117 & 0.118 & 0.119\\ 
(Q) V\{T,Q\}Z 2-body & 0.216 &&&&& 0.137\\
(Q) 3-body (a) & 0.018 &0.018&0.020&0.021&0.018 &0.018&0.018&0.017\\
(Q) 3-body (c) & 0.018 & 0.019&0.017&0.018&0.015&0.009&0.009&0.008\\
(Q) 4-body (a) & 0.015 &0.012&0.012&0.012&0.013&0.011&0.010&0.010\\
(Q) 5+6body (a) & -0.001 & 0.000 & 0.001 & 0.001 & 0.001 & 0.003 & 0.003 & 0.003\\
\hline
$T_4-(Q)$ 2-body VDZ(d,s)&-0.034&-0.034&-0.028&-0.028&-0.028&-0.021&-0.021&-0.021\\
$T_4-(Q)$ 2-body VTZ(d,p)&-0.029&-0.028&-0.020&-0.021&-0.022&-0.014&-0.014&-0.014\\
$T_4-(Q)$ 2-body VTZ(f,p)&-0.024&&&&&\\
$(5)$ 2-body VDZ(d,s)& 0.014 & 0.014 & 0.013 & 0.013 & 0.013 & 0.011 &0.011 &0.011\\\hline
CV 
2-body& 0.306 & 0.310 & 0.326 & 0.323 & 0.319 & 0.344 & 0.334 & 0.336\\
ditto 3-body & 0.009 & 0.008 & 0.013 & 0.010 & 0.011 & 
0.021 & 0.018 & 0.020\\
REL MP2/AVTZuc & -0.097 &
-0.093 &
-0.098 & 
-0.096 & 
-0.094 &
-0.106 &
-0.102 &
-0.103\\
ditto 3-body & -0.007 & -0.005& -0.008 & -0.006 &-0.008 & -0.014 & -0.012 & -0.014\\
ditto 4-body & 0.002 &0.002& 0.002 & 0.002&0.002&0.002&0.002&0.002\\
REL CCSD(T)/AVTZuc 2-body & -0.102&-0.098&-0.103&-0.101&-0.099&-0.111&-0.107&-0.109\\
DBOC (b) & 0.130 & 0.130 & 0.129 & 0.129 & 0.126 & 0.126 & 0.122 & 0.123  \\
\hline\hline
\end{tabular}
}

\begin{footnotesize}
(a) CCSDT(Q)/cc-pVDZ scaled\cite{jmlm200} by 1.25 as in W4lite;
 (b) CCSD/AVTZ 2-body; 3-body and 4-body terms vanish to all decimal places;
(c) CCSDT(Q)/cc-pVTZ(d,p) scaled\cite{jmlm200} by 1.10 as in W4 for cc-pVTZ.
\end{footnotesize}
\end{table}

As expected, the V\{D,T\}Z connected quadruples correction (Q) is binding overall. However, at least in the two-body term there is now substantial variation among the (H$_2$O)$_6$ isomers, from 0.19 kcal/mol for the prism to 0.12 kcal/mol for the ring in chair conformation. Decomposition into individual dimer contributions reveals nine nontrivial (and nearly equal) `pairs' for the prism --- corresponding to its nine vertices --- but (as stands to reason) only six for the ring.
For (H$_2$O)$_{20}$, only the 2-body terms were evaluated. Again we see variation between structures, from 0.70 kcal/mol for the dodecahedron to 0.85 kcal/mol for the box-kite structure. Like for (H$_2$O)$_{6}$, connected quadruples favor the more crowded structures over the more `open' ones.

Coming back to the hexamers: in order to verify basis set convergence, just for the prism and ring-boat structures, we recalculated the 2-body $T_3$-(T) and (Q) terms with the cc-pVQZ basis set and carried out V\{T,Q\}Z extrapolation. We obtained very similar results to the V\{D,T\}Z basis set pair, which comes with an about 20 times cheaper computational price tag.

The three-body contributions are an order of magnitude smaller than the two-body terms. With the smallest basis set, cc-pVDZ, all eight structures have similar contributions, but if that is increased to cc-pVTZ(d,p), we observe a two-way split between about 0.01 kcal/mol for the three ring structures and about 0.02 kcal/mol for the five remaining structures. The four-body term here is larger, 0.010 -- 0.015 kcal/mol, than for the higher-order triples, but still over an order of magnitude smaller than the dominant two-body term.

\begin{table}[h!]

\caption{Dissociation energy contributions for (H$_2$O)$_{20}$ Wales-Hodges isomers (kcal/mol)\label{tab:water20}}
\centering
\resizebox{\textwidth}{!}{
\begin{tabular}{lrrrr}\hline
& ES & FC & FS & \\
& edge-sharing & face[-sharing] & face-sharing & dodecahedron \\
& pentagons & cubes (box-kite) & stacked pentagons &\\
\hline
&\multicolumn{4}{c}{valence CCSD(F12*)/CBS +(T) limit}\\
Ref.\cite{jmlm272}& 219.30 & 216.31 & 217.58 & 211.22\\
Ref.\cite{jmlm294}& 219.19 & 215.98 & 217.03 & 211.58\\
\hline
$T_3-$(T) V\{D,T\}Z 2-body& -0.63 & -0.62 & -0.63 & -0.62 \\ 
$T_3-$(T) VDZ(d,s) 3-body& 0.077 & 0.091 & 0.079 & 0.059 \\
$T_3-$(T) VTZ(f,p) 3-body& 0.142 & 0.159 & 0.148 & 0.105 \\
$T_3-$(T) V\{D,T\}Z 3-body& 0.174 & 0.193 & 0.181 & 0.128 \\
(Q) 2-body & 0.81 & 0.85 & 0.83 & 0.70 \\
(Q) 3-body (a) & 0.098 & 0.081 & 0.084 & 0.093\\ 
CV CCSD(T)/awCV\{T,Q\}Z & 1.34 & 1.25 & 1.31 & 1.36\\
scalar relativistics 2-body & -0.41 & -0.39 & -0.40 & -0.38\\
DBOC & 0.57 & 0.54 & 0.56 & 0.55\\
\hline\hline
\end{tabular}
}

\begin{footnotesize}
Reference geometries taken unaltered from WATER27 dataset, via GMTKN55; 
(a) CCSDT(Q)/cc-pVDZ scaled\cite{jmlm200} by 1.25 as in W4lite.
\end{footnotesize}

\end{table}

As seen in general thermochemistry\cite{jmlm200,jmlm205}, the higher-order connected quadruples term $T_4 - (Q)$, i.e. the difference between fully iterative CCSDTQ\cite{Kucharski1992} and CCSDT(Q), is slightly antibonding. However, with the small cc-pVDZ basis set, this correction is somewhat exaggerated, and upon basis set expansion it drops down from --0.034 to --0.024 kcal/mol for the (H$_2$O)$_6$ prism, and --0.021 to --0.014 kcal/mol for the ring. Inclusion of connected quintuples,\cite{mrcc63,mrcc65} as shown previously\cite{jmlm200,jmlm205} for general thermochemistry, partly cancels this contribution, being binding by 0.014 and 0.011 kcal/mol, respectively, for prism and ring. (For this step, we used the cc-pVDZ(d,s) basis set, which was previously shown\cite{jmlm200,jmlm205} to be more than adequate for this high excitation contribution.) As there will be further cancellation of post-CCSDT(Q) terms with 3-body (Q) terms, we deem it both expedient and acceptable for the (H$_2$O)$_{20}$ isomers to ignore post-CCSDT(Q) correlation corrections.

For molecules without significant static correlation, the fairly systematic mutual cancellation between bonding quadruples and largely antibonding higher-order triples is what causes CCSD(T) to be a `Pauling point' (see also Stanton\cite{Sta97} for a perspective based on perturbation theory). This cancellation is broadly conserved here as well, though the considerable variation of the quadruples term between water hexamer structures causes small --- but in high-accuracy work significant --- differences between isomers. However, if we {\em do} concern ourselves with such small differences, it behooves us to investigate other potential sources of differences.

Scalar relativistic corrections reduce the dissociation energies of the (H$_2$O)$_6$ isomers across the board by about 0.1 kcal/mol, and those of the four (H$_2$O)$_{20}$ isomers by about 0.4 kcal/mol. This is fortuitously almost exactly compensated by the diagonal Born-Oppenheimer corrections, which {\em increase} the dissociation energies across the board by about 0.12--0.13 kcal/mol for (H$_2$O)$_6$ and by 0.56$\pm$0.01 kcal/mol for (H$_2$O)$_{20}$. The 3-body contribution to the scalar relativistic correction is an order of magnitude less than the two-body term, while the four-body term is altogether insignificant. The 3-body and 4-body DBOC terms vanish entirely.

The core-valence correlation is another matter. Its (bonding) two-body contribution reaches 0.30--0.34 kcal/mol for the (H$_2$O)$_6$ isomers, and 1.25--1.36 kcal/mol for the (H$_2$O)$_{20}$ isomers, thus becoming {\em the} most important term beyond valence CCSD(T) at the complete basis set limit.

It is generally assumed that core-valence (CV) correlation does not matter for noncovalent interactions. While we have evaluated it in high-accuracy work on benzene dimer and benzene...ethylene,\cite{jmlm298} it was found to be insignificant there. It would stand to reason that this would be the case for distant contacts; however, for closer contacts like hydrogen bonds, CV correlation can apparently not be ignored. Boese\cite{Boese2013} and \v{R}ez\'a\v{c}, Hobza, and coworkers\cite{A24bis} previously found 0.03 kcal/mol for water dimer: the values found presently are comparable to, if somewhat larger than, what one might expect from scaling this contribution by the number of hydrogen-bond contacts. Three-body effects are quite insignificant for core-valence correlation: they reach only 0.009--0.020 kcal/mol for the (H$_2$O)$_6$ isomers, or between one-fifteenth and one-thirtieth of the two-body terms. 

Overall, it is clear that the bulk  (90\% or more) of contributions beyond valence CCSD(T) are recovered by only 2-body terms. This is what makes their consideration for larger water clusters a viable proposition at all.

\section{Conclusions}

While for water clusters, post-CCSD(T) valence correlation contributions largely cancel between antibonding higher-order triples and connected quadruples, the cancellation is neither perfect nor homogenous: connected quadruples favor more compact structures (such as the water hexamer prism) over more spread-out ones (particularly rings). Overall, post-CCSD(T) valence correlation enhances the binding energy, and that is in fact reinforced when three-body terms are taken into account.

Relativistic corrections and DBOC fortuitously largely cancel each other. The most significant contribution for water clusters beyond  valence CCSD(T) is actually core-valence correlation.

Aside from higher order triple excitations, about nine-tenths of all these effects are recovered in the 2-body approximation, which makes their evaluation realistic also for larger clusters.

\begin{acknowledgement}
Research at Weizmann was supported by the Israel Science Foundation (grant 1969/20), by the Minerva Foundation (grant 2020/05), and 
by a research grant from the Artificial Intelligence for Smart Materials Research Fund, in Memory of Dr. Uriel Arnon. ES thanks the Feinberg Graduate School (Weizmann Institute of Science) for a doctoral fellowship and the Onassis Foundation (Scholarship ID: FZP 052-2/2021-2022).

This research was directly inspired by discussions with Prof. Kieron Burke (UC Irvine) about the (H$_2$O)$_6$ isomers. We additionally thank Profs. A. Daniel Boese (University of Graz, Austria) and Leslie Leiserowitz (Weizmann Institute) for helpful discussions, and Profs. Devin A. Matthews (Southern Methodist University, Dallas, Texas) 
for assistance with CFOUR and NCC.
\end{acknowledgement}

\section*{Data availability statement}

Raw data can be obtained from the senior author upon reasonable request.

\bibliography{waterNccsdtq}%
\end{document}